\documentclass[12pt]{article}
\usepackage{epsfig}
\def\be{\begin{equation}}
\def\ee{\end{equation}}
\newcommand{\bea}{\begin{eqnarray}}
\newcommand{\eea}{\end{eqnarray}}
\newcommand{\dst}{\displaystyle}
\newcommand{\fr}[2]{\frac{{\dst #1}}{{\dst #2}}}

\renewcommand{\Im}{{\rm Im}}
\date{\empty}

\newcommand{\f}{\phi}
\newcommand{\fd}{\phi^\dagger}

\def\lsim{\mathrel{\rlap{\lower4pt\hbox{\hskip1pt$\sim$}}
    \raise1pt\hbox{$<$}}}         %less than or approx. symbol

\def\gsim{\mathrel{\rlap{\lower4pt\hbox{\hskip1pt$\sim$}}
    \raise1pt\hbox{$>$}}}         %greater than or approx. symbol

\title{Two-Higgs-doublet model from the group-theoretic perspective}

\author{I.P. Ivanov\thanks{E-mail: igivanov@cs.infn.it}\\
  {\normalsize INFN, Cosenza, Italy}\\
  {\small and}\\
  {\normalsize Sobolev Institute of Mathematics,  Novosibirsk, Russia}}

\begin{document}
\maketitle

\begin{abstract}
In the two-Higgs-doublet model, different Higgs doublets can be viewed 
as components of a generic "hyperspinor".
We decompose the Higgs potential of this model into
irreducible representations of the $SU(2)$ 
group of transformations of this hyperspinor. We discuss 
invariant combinations of Higgs potential parameters $\lambda_i$
that arise in this decomposition and
provide simple and concise sets of conditions for the hidden $Z_2$-symmetry, 
Peccei-Quinn symmetry, and explicit $CP$-conservation in 2HDM.
We show that some results obtained previously by brute-force calculations 
are reduced to simple linear algebraic statements in our approach.
\end{abstract}

\section{Introduction}

The Electroweak Symmetry Breaking in the Standard Model is
described usually with the Higgs mechanism. Its simplest
realization is based on a single weak isodoublet of scalar fields,
which couple to the gauge and matter fields and self-interact
via the quartic potential, for review see \cite{Hunter}. 
Extended versions of the Higgs mechanisms are based on
more elaborate Higgs sectors. The simplest extension is known 
as the two-Higgs-doublet model (2HDM), which makes use 
of two Higgs weak isodoublets of scalar fields $\phi_1$
and $\phi_2$. This model has been extensively studied in literature from
various points of view, see \cite{Hunter,Sanchez,CP,ginzreview} and references
therein.

The Higgs potential $V_H = V_2 + V_4$ contains quadratic and quartic 
parts, which in the most general case of the 2HDM are conventionally parametrized as
\bea
V_2&=&-{1\over 2}\left[m_{11}^2(\phi_1^\dagger\phi_1) + m_{22}^2(\phi_2^\dagger\phi_2)
+ m_{12}^2 (\phi_1^\dagger\phi_2) + m_{12}^{2\ *} (\phi_2^\dagger\phi_1)\right]\,;\label{potential2}\\
V_4&=&\fr{\lambda_1}{2}(\phi_1^\dagger\phi_1)^2
+\fr{\lambda_2}{2}(\phi_2^\dagger\phi_2)^2
+\lambda_3(\phi_1^\dagger\phi_1) (\phi_2^\dagger\phi_2)
+\lambda_4(\phi_1^\dagger\phi_2) (\phi_2^\dagger\phi_1) \label{potential}\\
&+&\fr{1}{2}\left[\lambda_5(\phi_1^\dagger\phi_2)^2+
\lambda_5^*(\phi_2^\dagger\phi_1)^2\right]
+\left\{\left[\lambda_6(\phi_1^\dagger\phi_1)+\lambda_7
(\phi_2^\dagger\phi_2)\right](\phi_1^\dagger\phi_2) +{\rm
h.c.}\right\}\nonumber
\eea
with 14 free parameters: real $m_{11}^2, m_{22}^2, \lambda_1, \lambda_2, \lambda_3, \lambda_4$ 
and complex $m_{12}^2, \lambda_5, \lambda_6, \lambda_7$.
Note that in the case $m_{12}^2 = \lambda_6 = \lambda_7 = 0$ the potential
remains invariant under transformations from the group $Z_2$:
\be
\phi_1 \to -\phi_1\,, \phi_2 \to \phi_2\,,\quad\mbox{or}\quad 
\phi_1 \to \phi_1\,, \phi_2 \to -\phi_2\,.\label{z2}  
\ee
The full Higgs lagrangian includes also kinetic terms, which can be off-diagonal
in a general case, see discussion in \cite{ginzreview}. 
Having written lagrangian, one usually proceeds with calculating
physical observables in terms of this particular set of parameters.

Since the two Higgs doublets
$\phi_1$ and $\phi_2$ have the same quantum numbers, they can be viewed as
two components of a generic "hyperspinor".
Unitary rotations that mix these two doublets leave the structure of the 2HDM 
potential and --- more importantly --- the physical observables unchanged, 
modifying only the parameters of the lagrangian.
This freedom of choosing the basis for the two scalar doublets,
known as the Higgs basis or reparametrization invariance, was exploited 
in discussion of symmetries and invariants of the 2HDM 
\cite{ginzreview,lavoura1994,haber1}, 
$CP$-violation in the Higgs sector \cite{ginzreview,CPOddinv,haber2}, 
perturbative unitarity conditions \cite{gi2003}.

The importance of representing the lagrangian in a reparametrization-invariant way
was appreciated as early as in 1974, \cite{fayet}, in the context of supersymmetric theories.
In non-minimal SUSY, the two Higgs doublets were in fact members of a single Fayet-Sohnius hypermultiplet
related with each other by the ``flavor-$SU(2)$'' symmetry, \cite{fayet2}.

Recently, a general reformulation
the entire theory of the most general 2HDM in an explicitly $SU(2)$-covariant
form was presented in \cite{haber1}. 
The Higgs potential was written, following \cite{CP}, in a compact form as
\be
V_H = Y_{ab}(\fd_a \f_b) + Z_{abcd}(\fd_a \f_b)(\fd_c \f_d)\,,\quad \mbox{with}\quad
a,b = 1,2\,,
\label{yz}
\ee
and tensors of various ranks were constructed from $Y_{ab}$ and $Z_{abcd}$.
By contracting all the subscripts in different ways, a number
of quantities invariant under $SU(2)$ transformations were derived.
The physical meaning of these invariants, however, did not appear to be particularly clear.
Some specific choices of the Higgs potential parameters
were found to be important for the overall structure
of the 2HDM \cite{haber1} as well as for its $CP$-properties \cite{CPOddinv,haber2},
however their origin and interpretation were lacking.

In this note we present a simple yet revealing group-theoretic study
of the basis invariance in 2HDM. We identify the quartic Higgs potential as
a $C(2,2)$-tensor and decompose it in the sum of irreducible representations 
(irreps) of $SU(2)$. Several invariant combinations of $\lambda_i$ with
transparent origin follow immediately from this decomposition. 
Not intending to give an exhaustive list of all possible conclusions, we simply
demonstrate that the group-theoretic point of view proves useful
in understanding properties of 2HDM. 
We discuss in particular the conditions for the hidden $Z_2$-symmetry and
for the $CP$-conservation in 2HDM.
We show that many previous results of the brute-force calculations 
acquire an elegant interpretation in our approach.

\section{Decomposing quartic potential into irreps}

We consider the two scalar doublets $\phi_1$ and $\phi_2$ 
as two components of a "hyperspinor" $\psi_\alpha = (\phi_1,\ \phi_2)^T$, abstracting
from the "internal" structure of these doublets. 
The most general rotations of this spinor are operators from $SU(2)\times U(1)$
\begin{equation}
\left(\begin{array}{c}\phi_1' \\ \phi_2'\end{array}\right) =
e^{i\rho_0}\left(\begin{array}{cc}
e^{i\rho_1}\cos\theta  & e^{i\rho_2}\sin\theta  \\ 
- e^{-i\rho_2} \sin\theta & e^{-i\rho_1}\cos\theta \end{array}\right)
\left(\begin{array}{c}\phi_1 \\ \phi_2\end{array}\right)\,,
\label{su2}
\end{equation}
which is parametrized by $3+1$ independent parameters 
$\theta,\rho_1,\rho_2$ and $\rho_0$.
A particular case of this transformation $\rho_1 = \rho_0$, $\theta=0$  was introduced
in \cite{gv2001} as ``rephasing transformation'' and was used to remove the imaginary part
of the parameter $\lambda_5$. In \cite{gi2003} it was exploited in derivation 
of perturbative unitarity conditions for $CP$-violating 
case from those of the $CP$-conserving one.

The corresponding "antispinor" $\tilde\psi^\beta = (\phi_1^\dagger,\ \phi_2^\dagger)$ 
transforms under the antifundamental representation of $SU(2)$. 
With the aid of the antisymmetric tensor $\epsilon_{\gamma\beta}$, it can be mapped to
usual contravariant spinor $\tilde\psi_\gamma = \epsilon_{\gamma\beta}\tilde\psi^\beta
 = (\phi_2,\ -\phi_1)^T$, which transforms under the fundamental representation.

Let us focus on the quartic Higgs potential (\ref{potential}).
Being a quantity constructed from
two $\phi_i$ and two $\phi_i^\dagger$, it corresponds to a vector in the 16-dimensional 
complex space $C(2,2)$ spanned on the basis tensors
$\Psi_{\alpha_1 \alpha_2}^{\beta_1 \beta_2}$ with $\alpha_i,\beta_i = 1,2$,
constructed from various combinations of $\phi_i$ and $\phi_i^\dagger$,
or, alternatively, to a vector in the isomorphic space $C(4,0)$ spanned on
the basis vectorss 
\be
\Psi_{\gamma_1 \gamma_2 \alpha_1 \alpha_2} = 
\epsilon_{\gamma_1 \beta_1}\epsilon_{\gamma_2 \beta_2}
\Psi_{\alpha_1 \alpha_2}^{\beta_1 \beta_2}\,.\label{psi}
\ee
Since terms like $(\phi_1^\dagger \phi_2)(\phi_1^\dagger \phi_1)$ and
$(\phi_1^\dagger \phi_1)(\phi_1^\dagger \phi_2)$ represent the same pieces
of the potential, we in fact work
not in the full 16-dimensional space, but in its 10-dimentional subspace
symmetric under $(\alpha_1,\beta_1) \leftrightarrow (\alpha_2,\beta_2)$.
We will label this subspace with the subscript ``sym''. 

The coordinates of this vector either in $C(2,2)_{sym}$ or in $C(4,0)_{sym}$
are combinations of the parameters $\lambda_i$. Their transformation 
under rotations (\ref{su2}) of fields $\phi_i$ and, correspondingly,  
of basis vectors (\ref{psi}) defines 
a $C(2,2)_{sym}$ or in $C(4,0)_{sym}$ tensor (essentially the same tensor
as $Z_{abcd}$ in (\ref{yz})). 
The space of all possible parametrizations of the quartic Higgs potential 
realizes, therefore, the $C(2,2)_{sym}$ [or $C(4,0)_{sym}$] 
representation of the $SU(2)$ group.
One easily find its decomposition into irreps
\begin{equation}
(2\otimes 2\otimes 2\otimes 2)_{sym} = 5 \oplus 3 \oplus 1 \oplus 1\,.
\label{decompsymm}
\end{equation}
For a transparent derivation of this result, one merges first 
$\alpha_1\alpha_2$: $2\otimes 2 = 3 \oplus 1$,
where $3$ is symmetric and $1$ is antisymmetric under $\alpha_1\leftrightarrow \alpha_2$.
The same holds for $\gamma_1\gamma_2$. 
Their product gives $\left((3 \oplus 1)\otimes (3 \oplus 1)\right)_{sym} = 
3\otimes 3 \oplus 1\otimes 1 = 5 \oplus 3 \oplus 1 \oplus 1$.
In the conventional notation, the basis vectors in the space $C(2,2)_{sym}$ 
that form these multiplets are
\bea
\mbox{5-plet} &&
\begin{array}{lll}
|2,+2\rangle&=& (\fd_2\f_1)(\fd_2 \f_1)\,,\\[1mm]
|2,+1\rangle&=& (\fd_2\f_1)(\fd_2 \f_2) - (\fd_2\f_1)(\fd_1 \f_1) \,,\\[1mm]
|2,0\rangle&=& {1 \over\sqrt{6}}\left[(\fd_1\f_1)^2 + (\fd_2 \f_2)^2
- 2(\fd_1\f_1)(\fd_2\f_2) -  2(\fd_1\f_2)(\fd_2\f_1)\right]\,,\\[1mm]
|2,-1\rangle&=&  (\fd_1\f_2)(\fd_1 \f_1) - (\fd_1\f_2)(\fd_2 \f_2)\,,\\[1mm]
|2,-2\rangle&=& (\fd_1\f_2)(\fd_1 \f_2)\,,\\[1mm]
\end{array}\label{5plet}\\[1mm]
\mbox{3-plet} &&
\begin{array}{lll}
|1,+1\rangle&=& (\fd_2\f_1)\left[(\fd_1 \f_1) + (\fd_2 \f_2)\right] \,,\\[1mm]
|1,0\rangle&=& {1 \over\sqrt{2}}\left[(\fd_2\f_2) - (\fd_1 \f_1)\right]
\left[(\fd_1\f_1) + (\fd_2 \f_2)\right]\,,\\[1mm]
|1,-1\rangle&=& - (\fd_1\f_2)\left[(\fd_1 \f_1) + (\fd_2 \f_2)\right]\,,\\[1mm]
\end{array}\label{3plet}\\[1mm]
\mbox{singlet}_1 &&
\begin{array}{lll}
|0_1,0\rangle&=& {1 \over 2}\left[(\fd_1\f_1) + (\fd_2 \f_2)\right]^2\,,
\end{array}\label{singleta}\\[1mm]
\mbox{singlet}_2 &&
\begin{array}{lll}
|0_2,0\rangle&=& {1 \over \sqrt{12}}\left[\left((\fd_1\f_1) - (\fd_2 \f_2)\right)^2
+ 4(\fd_2\f_1)(\fd_1\f_2)\right]\,.
\end{array}\label{singletb}
\eea
Decomposition of the quartic potential (\ref{potential}) into these multiplets
\begin{equation}
V_4 = \sum_m a_{2,m} \cdot |2,m\rangle + \sum_n b_{1,n} \cdot |1,n\rangle
+ c |0_1,0\rangle + d |0_2,0\rangle\label{plets}
\end{equation}
yields the following coefficients
\bea
&&a_{2,+2} = \fr{\lambda_5^*}{2}\,,\quad
a_{2,+1} = - \fr{\lambda_6^* - \lambda_7^*}{2}\,, \quad
a_{2,0} = {\lambda_1 + \lambda_2 - 2\lambda_3 - 2\lambda_4 \over \sqrt{24}}\,,\nonumber\\[3mm]
&&a_{2,-2} = \fr{\lambda_5}{2}\,,\quad
a_{2,-1} = \fr{\lambda_6 - \lambda_7}{2}\,,\nonumber\\[3mm]
&&b_{1,+1} = {\lambda_6^* +\lambda_7^*\over 2}\,,\quad
b_{1,0} = - {\lambda_1 - \lambda_2 \over 2\sqrt{2}}\,,\quad
b_{1,-1} = -{\lambda_6 +\lambda_7\over 2}\nonumber\\[3mm]
&&c = {\lambda_1 + \lambda_2 + 2\lambda_3 \over 4}\,\quad
d = {\lambda_1 + \lambda_2 - 2\lambda_3 + 4\lambda_4  \over \sqrt{48}}
\eea
It is also convenient to exploit the homomorphism from $SU(2)$ to
$SO(3)$, switching from the triplet $b_{1,n}$
to a real vector $\vec b = (b_x, b_y, b_z)$ with
\begin{equation}
b_x = -\fr{1}{\sqrt{2}}\mbox{Re}(\lambda_6 + \lambda_7)\,,\quad
b_y = -\fr{1}{\sqrt{2}}\mbox{Im}(\lambda_6 + \lambda_7)\,,\quad
b_z = -\fr{1}{2\sqrt{2}}(\lambda_1 - \lambda_2)\,.\label{bi}
\end{equation}
Corresponding transformation of the 5-plet will turn it into
a real traceless tensor $a_{ij}$ ($i,j = x,y,z$)
\be
a_{ij} = {1 \over 2}
\left(\begin{array}{ccc}
\mbox{Re}\lambda_5 - a & \mbox{Im}\lambda_5 & 
\mbox{Re}(\lambda_6 - \lambda_7) \\
\mbox{Im}\lambda_5 & -\mbox{Re}\lambda_5 - a & 
 \mbox{Im}(\lambda_6 - \lambda_7) \\
\mbox{Re}(\lambda_6 - \lambda_7) & \mbox{Im}(\lambda_6 - \lambda_7) & 2a
\end{array}\right)\,,\label{aij}
\end{equation}
where $a \equiv (\lambda_1 + \lambda_2 - 2\lambda_3 - 2 \lambda_4)/6$.

The quadratic part $V_2$ of the Higgs potential (\ref{potential2})
can also be decomposed into irreps    
\be
V_2 =  \sum_n y_{1,n} \cdot |1,n\rangle + f |0_3,0\rangle \label{plets2}
\ee
with
\begin{equation}
y_{1,+1} = -{m_{12}^{2\ *} \over 2}\,,\quad
y_{1,0} = {m_{11}^2 - m_{22}^2 \over 2\sqrt{2}}\,,\quad
y_{1,-1} = {m_{12}^2 \over 2}\,,
\quad f = - {m_{11}^2 + m_{22}^2 \over 2\sqrt{2}}\,,
\end{equation}
or, analogously to (\ref{bi}), 
\begin{equation}
y_x = \fr{1}{\sqrt{2}}\mbox{Re}\, m_{12}^2\,,\quad
y_y = \fr{1}{\sqrt{2}}\mbox{Im}\, m_{12}^2\,,\quad
y_z = \fr{1}{2\sqrt{2}}(m_{11}^2 - m_{22}^2)\,.\label{yi}
\end{equation}

\section{Consequences}

\subsection{Invariants}
Several combinations of $\lambda_i$
remain invariant under an arbitrary $SU(2)$ rotation.
Our decomposition immediately reveals
four of them related to weights
of each irrep in (\ref{plets}). Two of them are linear 
and the other two are quadratic combinations of $\lambda_i$: 
\bea
\mbox{singlet}_1: & & \lambda_1 + \lambda_2 + 2\lambda_3 = \mbox{ const}\,,\nonumber\\
\mbox{singlet}_2: & & \lambda_1 + \lambda_2 - 2\lambda_3 + 4\lambda_4 
= \mbox{ const}\,,\nonumber\\
\mbox{3-plet}: & & (\lambda_1 - \lambda_2)^2 + 4|\lambda_6 + \lambda_7|^2 
= \mbox{ const}\,,\label{invariants}\\
\mbox{5-plet}: & & |\lambda_5|^2 + |\lambda_6 - \lambda_7|^2 
+ {1\over 12}(\lambda_1 + \lambda_2 - 2\lambda_3 - 2\lambda_4)^2
= \mbox{ const}\,.\nonumber
\eea
In total, there are 7 algebraically independent invariant combinations of $\lambda_i$ only.
This is not surprising, since we have 10 parameters of the Higgs potential and 3 degrees of freedom
in reparametrization transformations. 
By denoting $b^{(k)}_i \equiv (a^k)_{ij} b_j$, one can select the following independent invariants
\be  
c\,,\  d\,,\ \mbox{Tr}(a^2)\,, \ \mbox{Tr}(a^3)\,, \ \vec b^2\,,\ 
(\vec b \vec b^{(1)})\,, \ \epsilon_{ijk}\, b_i\, b_{j}^{(1)}\, b_{k}^{(2)}\,,
\label{otherinvs} 
\ee
where the last combination is just the 
scalar triple product of vectors $\vec b$, $\vec b^{(1)}$, and $\vec b^{(2)}$.

Any other algebraic function of $\lambda_i$ invariant under a generic reparametrization
transformation can be expressed as an algebraic function (and not always a polynomial) 
of invariants (\ref{otherinvs}).
The proof consists in straightforward application of linear algebra to our problem.
Consider first invariant combinations constructed from the matrix $a_{ij}$ only.
They are built of scalars Tr$(a^k) = x_1^k + x_2^k + x_3^k$, where $x_i$ are the eigenvalues of 
matrix $a_{ij}$. Since the characteristic polynomial for $a_{ij}$
has coefficients proportional to Tr$(a^2)$ and Tr$(a^3)$, and since any symmetric 
polynomial of $x_1$, $x_2$, $x_3$ can be written as a polynomial of 
$x_1 + x_2 + x_3 = 0$, $x_1x_2 + x_2x_3 + x_3x_1 = -\mbox{Tr}(a^2)/2$, and $x_1 x_2 x_3 = \mbox{Tr}(a^3)/3$, 
it follows that Tr$(a^k)$ for $k>3$ is always a polynomial of Tr$(a^2)$ and Tr$(a^3)$.

Now consider invariants that contain $a_{ij}$ as well vector $\vec b$.
There are at most three linearly independent vectors, which can be chosen $\vec b$, $\vec b^{(1)}$, and their
cross-product $[\vec b, \vec b^{(1)}]$. Any other $b^{(k)}_i$ is expressible as a linear combination 
of these three. On passing to scalar invariants, one recovers the last three expressions
in (\ref{otherinvs}).

The way the invariants (\ref{invariants}) are derived offers 
a transparent interpretation of some results of direct calculations.
For example, the authors of \cite{haber1}, after inspection of some reduced tensors,
note that if $\lambda_1 = \lambda_2, \lambda_6 = -\lambda_7$ 
holds in one basis, it will also hold after an arbitrary $SU(2)$ rotation. 
In our approach, this follows immediately from 
the fact that absence of the triplet in decomposition (\ref{plets}) 
in a basis independent statement.

If one takes into account also the quadratic part (\ref{potential2}) of the Higgs potential, a number
of other invariants arises. They include (\ref{otherinvs}) with $\vec b \to \vec y$
and additional mixed invariants that involve both $\vec b$ and $\vec y$.
Writing down all of them would be a lengthy but rather straightforward excercise.
It offers a much simpler procedure to listing all independent invariants than
the number-crunching approach of \cite{haber2}.

\subsection{Hidden $Z_2$-symmetry}
The specific case $m_{12}^2 = \lambda_6 = \lambda_7 = 0$ is of particular interest in 2HDM, since
in this case the $Z_2$-symmetry (\ref{z2}) of the Higgs potential is restored.
In the case  $\lambda_6 = \lambda_7 = 0$, but $m_{12}^2 \not = 0$ one speaks about
soft $Z_2$-violation. 

In the case of explicitly broken $Z_2$ symmetry by presence of $m_{12}^2$, $\lambda_6$ or 
$\lambda_7$, it might be still possible that there exists some $SU(2)$ rotation which
sets these coefficients to zero (the hidden $Z_2$ symmetry).
A question then arises: under what condition a general 2HDM potential
possesses this hidden $Z_2$ symmetry. 
Of course, a brute-force calculation 
(start from (\ref{potential}), perform $SU(2)$ rotation,
set resulting $\lambda_6'$ and $\lambda_7'$ to zero) 
can yield those conditions in an explicit algebraic form, 
for necessary formulae see \cite{ginzreview,haber1,haber2}.  
However, the group theoretical analysis performed above
offers an elegant formulation of these conditions.

Setting $\lambda_6 = \lambda_7 =0$ requires
removing {\em simultaneously} $\lambda_6 + \lambda_7$ from vector $\vec b$ (\ref{bi})
(by choosing $z$ axis along $\vec b$) and $\lambda_6 - \lambda_7$ from tensor 
$a_{ij}$ (\ref{aij}), which amounts to bringing tensor to its principal axes
(recall that once $\lambda_6=\lambda_7=0$, simple rephasing transformation 
removes Im$\lambda_5$). Removing $m_{12}^2$ means that, in addition, $\vec y\, \|\, \vec b$.
All this is possible if and only if 
the direction of vector $\vec b\, \|\, \vec y$ coincides with one of the principal 
axes of the tensor $a_{ij}$. Put in algebraic terms, 
\be
\parbox[c]{13cm}{
hidden $Z_2$ symmetry holds if and only if vectors $\vec b$ and $\vec y$ are collinear and 
are eigenvectors of $a_{ij}$.}
\label{Z2}
\ee
Degenerate cases like
absence of vector or tensor in decomposition (\ref{plets}) are also included.

Some choices of parametrization considered in 
\cite{haber1,haber2} again have transparent meaning in our approach.
In particular, the quartic potential was shown there 
to possess the hidden $Z_2$-symmetry, in particular,
in two cases: when $\lambda_1 = \lambda_2, \lambda_6 = -\lambda_7$ 
or when $\lambda_1+\lambda_2 = 2\lambda_3+2\lambda_4$, 
$\lambda_5 = 0$ and $\lambda_6 = \lambda_7$. 
In our language this conclusion immediately follows 
from the absence of triplet or 5-plet in (\ref{plets}), 
which automatically fulfils the above requirement. 

\subsection{Global Peccei-Quinn $U(1)$-symmetry}
If the potential possesses the hidden $Z_2$ symmetry and if, in addition,
$\lambda_5$ happens to be zero in the basis where $m_{12}^2 = \lambda_6 = \lambda_7 = 0$,
the Higgs potential is said to possess the global Peccei-Quinn (PQ) $U(1)$-symmetry \cite{PQ},
although this situation was considered in \cite{fayet} even before the work of 
Peccei and Quinn in the context of supersymmetric theories.
A close inspection of the matrix $a_{ij}$ reveals
the reparametrization-invariant criterion of the existence of this symmetry:
\be
\parbox[c]{13cm}{the PQ symmetry holds, if and only if
two eigenvalues of matrix $a_{ij}$ coincide {\em and} vectors $\vec b$ and $\vec y$
are both eigenvectors of $a_{ij}$ corresponding to {\em the other}, third, egienvalue.} 
\label{PQ}
\ee
The latter condition in (\ref{PQ}) removes possibilities $|\lambda_5| = \pm 3a$, in which case
the spectrum of $a_{ij}$ is also degenerate, but no PQ symmetry is realized.

\subsection{Explicit $CP$-conservation}

The two-Higgs doublet model enjoys so much attention
because it provides room for $CP$-violation originating from the Higgs sector \cite{classics}.  
A neccessary and sufficient condition for the Higgs potential to explicitly conserve $CP$
is that all the coefficients in the Higgs potential be real, after an appropriate
reparametrization transformation (Theorem 1 in \cite{haber2}).

As can be seen directly from (\ref{bi}), (\ref{aij}), (\ref{yi}), 
condition $\Im\, m_{12}^2 = \Im \lambda_5 = \Im \lambda_6 = \Im \lambda_7 = 0$
means ``decoupling'' of the $y$-direction from the $x$ and $z$-directions.
Formulated in a reparametrization-invariant way, this means that
\be
\parbox[c]{13cm}{ the Higgs potential is explicitly $CP$-conserving if and only if 
there exists an eigenvector of $a_{ij}$ orthogonal to both $\vec b$ and $\vec y$.} 
\label{CP}
\ee
In a non-degenerate case, when both $\vec b$ and $\vec y$ are non-zero,
the above statement means that
the cross-product of the two vectors must be an eigenvector of $a_{ij}$:
\be
a_{ij}\epsilon_{jkl} b_k y_l \propto  \epsilon_{ikl} b_k y_l\,.
\label{cross}
\ee 
Contracting (\ref{cross}) with $b_i$ or $y_i$, one finds two scalar conditions
for $CP$-conservation:
\be
\epsilon_{jkl} a_{ij} b_i b_k y_l = 0 \quad \mbox{and} \quad
\epsilon_{jkl} a_{ij} y_i b_k y_l = 0 \,.
\label{2scalars}
\ee
In a degenerate case, when $\vec b = 0$ or $\vec y = 0$, we are left with
a half of the requirement (\ref{CP}). 
For example, if $\vec b = 0$, we require that $\vec y$ be orthogonal to some
of the eigenvectors of $a_{ij}$. This takes place
if and only if the triple scalar product
\be
[\vec y, \vec y^{(1)}, \vec y^{(2)}] = 0\,, \quad \mbox{where}\quad y^{(1)}_i \equiv a_{ij}y_j\,,\quad
y^{(2)}_i \equiv a_{ij}y^{(1)}_j\,,
\label{eee}
\ee
In the other degenerate case, when $\vec y = 0$, the condition for $CP$-conservation reads:
\be
[\vec b, \vec b^{(1)}, \vec b^{(2)}] = 0\,.
\label{eee2}
\ee
Both (\ref{eee}) and (\ref{eee2}) are obviously reparametrization-invariant conditions.

The condition (\ref{CP}) is arguably more compact and transparent than those 
published before.  
The reparametrization invariant conditions for the Higgs potential in 2HDM to be explicitly $CP$-conserving 
have been analyzed recently in \cite{CPOddinv} and \cite{haber2}.
The results of these two studies did not coincide:
the authors of \cite{CPOddinv} found three sufficient and necessary conditions
for the Higgs potential to be $CP$-conserving, while authors \cite{haber2}
list four independent conditions.

The two papers agree that for a non-degenerate case vanishing of two invariants, denoted $I_1$ and $I_2$
in \cite{CPOddinv} and $I_{Y3Z}$ and $I_{2Y2Z}$ in \cite{haber2}, are necessary and sufficient
for $CP$-conservation. 
It is for degenerate cases that the conclusions of these two papers differ.
Authors of \cite{CPOddinv} argued that for $m_1 = m_2$ the two invariants vanish, while
the possibility for the $CP$-violation still remains within the quartic part of the Higgs potential only.
In order to remove this possibility, they set to zero another invariant $I_3$ constructed 
from product of 6 tensors $Z_{abcd}$.

The authors of \cite{haber2} provide their counterpart of this invariant, $I_{6Z}$,
and proceed further to analyze another degenerate case $\lambda_1 = \lambda_2$, $\lambda_6 + \lambda_7 = 0$.
They argue that all three previous invariants vanish in this case, yet
the possibility for the $CP$-violation remains.
To eliminate it, one must set to zero a fourth invariant, $I_{3Y3Z}$.
They claim that setting to zero these four invariants is necessary and sufficient
for explicit $CP$-conservation in the Higgs sector.

Our study confirms the results of \cite{haber2}. We see that last degenerate case considered there
means in our notation $\vec b = 0$, therefore their fourth invariant must
represent the same condition as our (\ref{eee}).
The other three invariants also have clear counterparts in our notation.
A question, however, remains about the {\em minimal} set of conditions that would grasp all cases.
Clearly, (\ref{eee}) and (\ref{eee2}) alone are not sufficient, since they do not 
include the requirement that both vectors be orthogonal to {\em the same} eigenvector
of $a_{ij}$.

It appears that the flaw in the analysis of \cite{CPOddinv} was overlooking the possibility
$c_1 = c_2\,,\ \theta_2 = \theta_1 + \pi$, which amounts to $\lambda_6 + \lambda_7 = 0$ in the usual notation.
This leads to $I_1 = I_2 = 0$ without fixing the value of $\theta_1$.
Assuming, in addition, $a_1 = a_2$, one makes zero also the third invariant $I_3$,
still without fixing $\theta_1$. To remove this residual possibility for $CP$-violation,
one must impose another, the fourth, condition. 

In our analysis, we do not discuss {\em spontaneous} $CP$-violation, since
it requires knowledge of the hyperspinor of the vacuum expectation values.

\subsection{RG evolution of $\lambda_i$}
The approach proposed here can also help better understand the 
generic properties of the renormalization-group (RG)
evolution flows of quartic coupling constants $\lambda_i$ in 2HDM.
These equations were explicitly written in \cite{lavoura2004}
in a general tensor-like form.
In our approach, these are replaced by equations on matrix $a_{ij}$, 
on vector $b_i$ and on scalars $c$ and $d$.

Restoration of symmetries under the RG flows is a well known phenomenon, for instance, in $O(n)$ models, 
see \cite{book,ginzsymm}. One might expect similar
phenomena to happen in 2HDM. In order to see this from equations without solving them,
one can write these equations in a generic tensorial form as
\bea
{d a_{ij} \over dt} &=& \beta_{a} \left(a_{ik} a_{kj} - {1\over 3}\mbox{Tr}(a^2)\delta_{ij}\right) 
+ \beta_{b} \left(b_i b_j - {1\over 3}b^2\delta_{ij}\right) + \mbox{{\it trivial}}\,,
\label{RGa}\\
{d b_i \over dt} &=& \beta_{ab}\, a_{ij} b_j + \mbox{{\it trivial}}\,.
\label{RGb}
\eea
Here {\it trivial} indicates terms with trivial tensorial structure, {\it i.e.}
$\propto a_{ij}$ for (\ref{RGa}) and $\propto b_i$ for (\ref{RGb}).
We focus now on the relative orientation of the vector $\vec b$ and the principal axes
of $a_{ij}$. 
One sees that there is a $Z_2$-symmetric fixed point of these equations.
There are grounds to expect that this fixed point is {\it stable}.
Indeed, in a suitable gauge theory with a single $SU(N)$ field \cite{cheng} (instead of the 
electroweak $SU(2)\times U(1)$ structure), the asymptotically-free solution will have 
$$
a_{ij}(t) = {\bar{a}_{ij} \over t}\,, \quad b_i(t) = {\bar{b}_i \over t^k}\,,
$$
for some $k \ge 1$ to be determined from the equations, 
which turns (\ref{RGa}) and (\ref{RGb}) into algebraic equations.
In particular, (\ref{RGb}) after this substitution simply states that vector $\vec b$ is an eigenvector
of $a_{ij}$. We conclude that the $Z_2$ symmetry is restored in this particular theory 
in the scaling limit.

Although in the true electroweak theory such a simple scaling limit does not exist,
the tendency towards restoration of $Z_2$ symmetry still might take place.
Indeed, contraction of a real symmetric matrix with a real vector can be represented
as a double vector product $[\vec \omega [\vec \omega \vec b]]$ modulo to
diagonal terms. Thus, vector $\vec b$ is attracted towards one of the
principal axes of $a_{ij}$ under the RG evolution.
A more detailed analysis is needed to establish the symmetry dynamics under the RG flow.

\subsection{Extensions and generalizations}

In the above study we did not provide expressions for the vacuum expectation 
values of the 2HDM. This analysis requires further study, since in this
case the fundamental degree of freedom, the hyperspinor $\psi_\alpha$
splits into two daughter hyperspinors of the vacuum expectation values and
residual fields. The equation for the hyperspinor of vacuum expectation values was written
in \cite{haber1} without explicit solution.  On the other hand, the properties of vacuum in the 2HDM
have received recently some attention, \cite{vev}.
We expect that our group-theoretical analysis might be extended to embrace these
issues as well. Algebraic structures that should arise along these lines
require a closer look. 

Finally, we note that our approach is applicable to
more involved realizations of spontaneous breaking of the 
electroweak symmetry. For example, in a general $n$-Higgs doublet model 
(nHDM), the quartic potential has $n^2(n^2+1)/2$ terms 
(including mutually conjugate ones) giving rise to equally 
many real parameters $\lambda_i$.
The number of irreps in $C(2,2)_{sym}$
is 5 for $n=3$ and 6 for $n>3$, including two scalars.
Thus, for nHDM one can identify two invariants linear in $\lambda_i$,
3 (for $n=3$) or 4 (for $n>3$) invariants quadratic in $\lambda_i$,
and many further invariants of higher order, which can be obtained
by contracting various irreps. Although in this case the simplicity 
of the linear algebraic analysis of 2HDM case is lost,
one might still expect to gain some insight from the group-theoretic approach.

\section{Summary}

We presented a simple and transparent
group-theoretic interpretation of several phenomena in 2HDM.
We treated the quartic potential as a vector in $C(2,2)$ space
formed by the hyperspinor $(\phi_1,\phi_2)$ and decomposed it
into a sum of irreducible representations of the $SU(2)$ group:
two scalars, one triplet and one 5-plet. 
This helped us reduce lengthy tensor analysis of 2HDM to simple linear algebra.

Within our formalism, we discussed the meaning of some invariants,
which have been discovered previously with the aid of straightforward calculations.
We presented simple, explicit and reparametrization-invariant conditions
for $Z_2$-symmetry, Peccei-Quinn symmetry and $CP$-conservation of the Higgs
potential in 2HDM. We argue that further development of this approach 
both to 2HDM and more involved Higgs sectors should prove very interesting. \\

I am thankful to Ilya Ginzburg for numerous discussions
and to P.~Fayet, L.~Lavoura, E.~Vdovin, and the referee for useful comments
and suggestions.
This work was supported by the INFN Fellowship, 
and partly by INTAS and grants RFBR 05-02-16211 and NSh-2339.2003.2.

\end{document}